\documentclass[conference]{IEEEtran}
\IEEEoverridecommandlockouts
\usepackage{cite}
\usepackage{amsmath,amssymb,amsfonts}
\usepackage{algorithmic}
\usepackage{graphicx}
\usepackage{textcomp}
\usepackage{xcolor}
\usepackage{multirow}
\def\BibTeX{{\rm B\kern-.05em{\sc i\kern-.025em b}\kern-.08em
    T\kern-.1667em\lower.7ex\hbox{E}\kern-.125emX}}
\begin{document}

\title{Enhancement and Evaluation of MOXcatter\\
}

\author{\IEEEauthorblockN{Yucong Gao}
\IEEEauthorblockA{\textit{School of Computer Science} \\
\textit{University of Science and Technology of China}\\
}
}

\maketitle

\begin{abstract}
    Backscatter WiFi provides a novel solution to IoT device's energy consumption. Different from other backscatter WiFi solutions, MOXcatter works with multiple spatial streams, making it appliable in 802.11n and beyond. In this paper, we present a method to solve the problem of phase ambiguity in MOXcatter due to the deviation of packet offset measurement. The experiments show that by changing the modulation method from BPSK to DBPSK, MOXcatter achieves up to 10 dB gain in BER. We evaluate its performance and discuss possible enhancements and applications.
\end{abstract}


\section{Introduction}
The most common usage scenario for Internet of Things applications is probably indoor wireless environment. Being the most widely used technology in this scenario, 802.11 WiFi is always the prominent choice. Different form Bluetooth, the WiFi transceiver consumes more power. A typical WiFi transceiver working in 75 Bps consumes 80mW, which is 40x higher than Bluetooth\cite{b1}, making energy consumption a concern. \par
For WiFi and many other existing wireless devices, much of the energy are consumed while transmitting signals, but backscatter communication has changed the stereotype recently\cite{b2}. In a backscatter communication system, excitation signals are provided by a reader, and a tag device uses ASIC (Application Specific Integrated Circuit) or FPGA (Field Programmable Gateway Array) to modulate and reflect an excitation signal. After receiving the reflected signal, the reader can then decode the tag data accordingly. Since that the tag device doesn't have to proactively generate excitation signal, backscattered systems can be highly energy efficient. A tag can be completely battery-free if harvested energy from the excitation signals is sufficient for powering the computation and transmission units on board. There have been significant studies on backscatter communications, e.g., RFID (Radio Frequency IDentification), which is available in the mass market with very low cost\cite{b3}. Recent studies suggest that, backscatter systems can not only based on RFID signals, but also FM radio\cite{b4}, BLE (Bluetooth Low Energy) \cite{b5},ZigBee\cite{b6}, WiFi\cite{b7}, etc.\par
Since that many of the IoT devices are designed to be compatible with WiFi but not every WiFi functionality are needed, there are many studies based on WiFi backscatter.For example, passive Wi-Fi\cite{b8} has enhanced the application of backscatter systems to work with commercial WiFi devices, however requires additional device to generate the excitation signal. Inter-Technology Backscatter \cite{b9}, HitchHike \cite{b10}, and FreeRider \cite{b11} all work with off-the-shelf WiFi but only deal with single-antenna signals, making their performance constrained for MIMO (Multiple-Input Multiple-Output) -based signals, which has been the cornerstone for 802.11n and beyond. To explore the full potential of the latest WiFi standards, MOXcatter is introduced, which is a spatial-stream backscatter system using commodity multi-antenna WiFi\cite{b7}. The backscatter tag uses the spatial stream data to carry the tag information; a backscatter WiFi receiver, which is a commodity WiFi device can decodes the tag information by comparing the backscattered data with the original data received from another ambient WiFi receiver, which is also a commodity WiFi device.\par
The key contribution of this paper is the enhancement of MOXcatter system by the implementation of DPSK and the evaluation of DBPSK modulation working with 802.11n MIMO-OFDM (Orthogonal Frequency-Division Multiplexing) Single-Stream (SS) 6.5Mbps.

\section{MOXcatter Design}
In this section, we will prove the feasibility of modulating tag signal with phase change on an existing signal while modulated signal can still be a legit one. 

\subsection{Tag Codebook for Spatial Stream signal}
MOXcatter system is different from the previous backscatter WiFi solution since the tag information is conveyed in OFDM symbol level using phase modulation. The sequence of \textit{m} complex coded data contained in the \textit{n}th OFDM symbol can be formulated as follows:  
\begin{equation}
        {Y_n} = \left\{ {{y_1}^{\left( n \right)},{y_2}^{\left( n \right)},...,{y_m}^{\left( n \right)}} \right\}
\end{equation}
         where ${y_r}^{\left( n \right)}$ is the I/Q data on the subcarrier \textit{r} of symbol \textit{n}. ${\rm{Y}}=\left[Y_1,Y_2,...,Y_l\right]'$ is the vector of the transmitted OFDM symbols.\par

         MOXcatter follows a codebook to modulate tag through changing the subcarrier phase. This procedure can be represent by multiplying the OFDM symbol vector with the modulation vector $\Theta   =  [{\theta _1},{\theta _2},...,{\theta _l}]$, where ${\theta _i} \in \left\{ { - 1,1} \right\},i \in \left\{ {1,2,...,l} \right\}$ for subcarrier using BPSK, and ${\theta _i} \in \left\{ { - 1,1,{e^{ - j\frac{\pi }{2}}},{e^{j\frac{\pi }{2}}}} \right\},i \in \left\{ {1,2,...,l} \right\}$ for QPSK,16-QAM, 64-QAM and beyond. After multiplying, the tag-modulated OFDM symbol sequence can be expressed as follows:
\begin{equation}
    {{\rm{Y}}^{tag}} = \Theta  \cdot {\rm{Y}}
\end{equation}
Obviously, the tag modulation is a linear transform of the I/Q data.\par
\begin{figure}[htbp]
    \centerline{\includegraphics[width=1\linewidth]{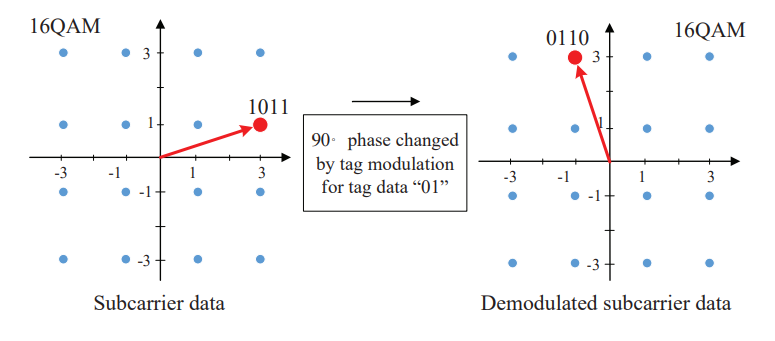}}
    \caption{An illustration of why the tag-modulated subcarrier data still fall in the constellation of QAM and hence can be demodulated by a commodity WiFi device\cite{b7}.}
    \label{fig1}
\end{figure}
\begin{table}[htbp]
    \caption{Delay Profile Model}
    \begin{center}
    \begin{tabular}{|c|c|c|}
    \hline 
    \textbf{Subcarrier} & \textbf{Subcarrier phase}& \multirow{2}{*}{\textbf{Tag data}} \\
    \textbf{modulation} & \textbf{changed by tag} & \\
    \hline
    \multirow{2}{*}{BPSK} & $0^{\circ}$ & 0 \\
    \cline{2-3}
    & $180^{\circ}$ & 1 \\
    \hline
    \multirow{4}{*}{QPSK,QAM} & $0^{\circ}$ & 00 \\
    \cline{2-3}
    & $90^{\circ}$ & 01 \\
    \cline{2-3}
    & $180^{\circ}$ & 10 \\
    \cline{2-3}
    & $270^{\circ}$ & 11 \\
    \hline
    \end{tabular}
    \label{tab1}
    \end{center}
    \end{table}
Study in \cite{b7} has proved that when tag modulation changes the phase of the OFDM symbol waveform, it actually changes the I/Q data. According to the constellation analysis in Figure 1 and the tag codebook in Table 1,original data and tag-modulated subcarrier data have the same codebook.

\subsection{Tag Modulation}
According to the analysis in the last Subsection, if an OFDM symbol's phase is changed by a MOXcatter tag following the codebook in Table 1, the backscattered signal can still be decoded by commodity WiFi devices. We simulate this modulation procedure by multiplying the original 802.11n signal with a square wave signal, whose phase changes when the tag data changes. This procedure does not change the phase of the preamble in the PLCP (PHY Layer Convergence Procedure), so as to ensure that each backscattered packet can be decoded properly. The symbol phase is then changed according to the tag data and the codebook.\par
MOXcatter tag applys a frequency shift to the 802.11n signal, so the receivers can distinguish the backscattered signal from the original one and keep the ongoing data communication unaffected.\par
\subsection{Tag Information Decoding}
The backscattered signal is received and decoded by a commodity WiFi device with multiple antennas, e.g., with an 802.11n NIC. As is mentioned in in the last Subsection, the original 802.11n packets and the backscattered 802.11n packets will be received at two different frequencies with a 
\begin{figure}[htbp]
    \centerline{\includegraphics[width=1\linewidth]{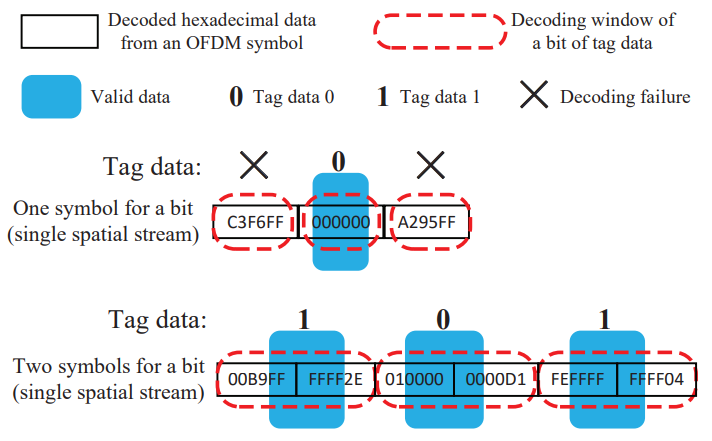}}
    \caption{Tag data decoding using a decoding window and a search for fixed-length all-zero (all-one) sequences\cite{b7}.}
    \label{fig2}
\end{figure}
\noindent fixed difference which can be controlled by the frequency of the square wave. These two packet will be used together to decode the tag information in two steps.\par

The first step is to performe an XOR operation on the original 802.11n packet and the backscattered 802.11n packet. Let ${\rm{\bf{Q}}} = \left\{Q_1,...,Q_n\right\}$ be the original 802.11n data, where $Q_i$ is the data bits of the $i$th OFDM symbol and ${\rm{\bf{B}}} = \left\{B_1,...,B_n\right\}$ be the backscattered 802.11n data, where $B_i$ is the data bits of the $i$th OFDM symbol. The processed raw data ${\rm{\bf{\Gamma}}} = \left\{\Gamma_1,...,\Gamma_n\right\}$ can be obtained as follows:
\begin{equation}
    {\Gamma _i} = {Q_i} \oplus {B_i}
\end{equation}
where $\Gamma_i$ is the processed raw data bits from the $i$th OFDM symbol. The length of $\Gamma_i$ depends on the modulation methods, e.g., $\left| {{\Gamma _i}} \right| = 24 $ for 802.11g OFDM BPSK subcarrier coding rate $\frac{1}{2}$, $\left| {{\Gamma _i}} \right| = 26 $ for 802.11n MIMO-OFDM single stream BPSK subcarrier. Since the tag information is carried by more than one bit, another step is required to decode the tag information in the OFDM-symbol level in our system.\par

In the second step, we need search and decode the specific sequence of
a fixed-length $\left| {{\Gamma _i}} \right|$ For BPSK or DBPSK subcarrier modulation, the sequence is all-zero (or all-one).In the tag modulation, we can choose the number of the OFDM symbols that are used to carry one bit tag data. If we use two OFDM symbols to carry tag data 1, the corresponding time-domain waveform of the two OFDM symbols will be changed by a 180 degree phase shift. Figure 2 is an illustrative example form \cite{b7}, where a tag transmits a data sequence “101" using a single spatial stream signal. For the BPSK-subcarrier modulation used for 802.11n, the tag data decoding may fail when we use one symbol for one bit of the tag data; using a decoding window of length $2\left| {{\Gamma _i}} \right|$, the tag data can be successfully decoded. Our simulation also shows the same phenomenon.

\section{Enhancement Of MOXcatter}
Through the introduction of MOXcatter, we have proved the feasibility of modulating tag signal on OFDM symbols. However, in the real multipath fading channel, demodulation of the tag information will be disrupted by many factor.\par

\begin{figure}[htbp]
    \centerline{\includegraphics[width=1\linewidth]{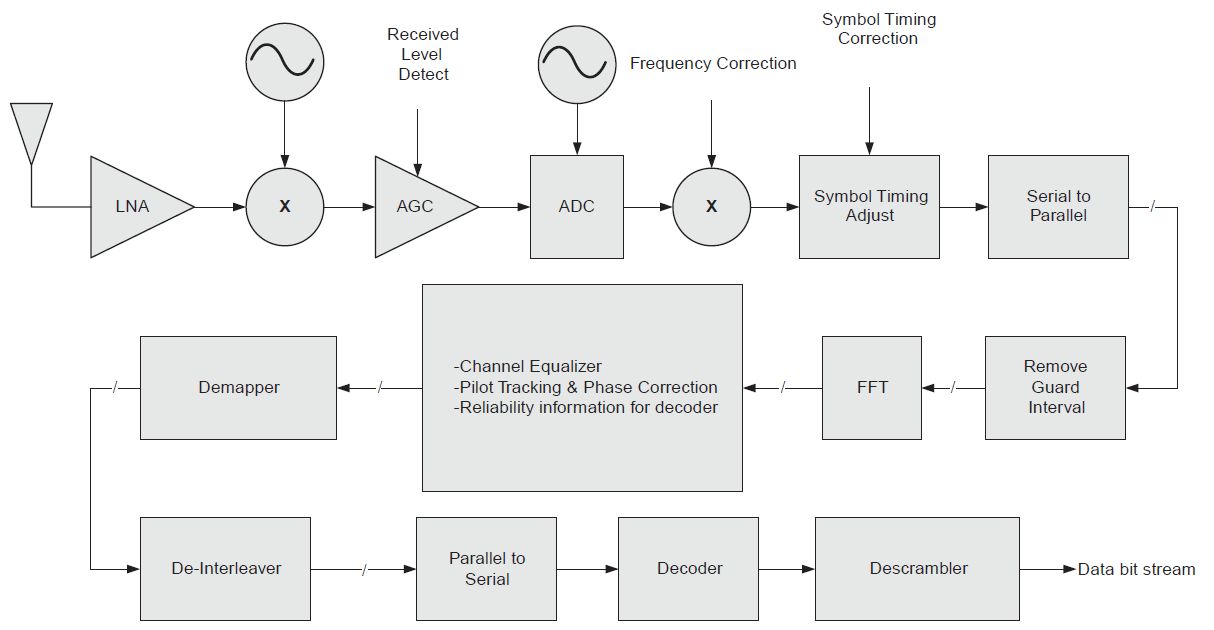}}
    \caption{Typical PHY receive procedure of OFDM-based 802.11 protocols\cite{b12}.}
    \label{fig3}
\end{figure}

Figure 3 shows the typical PHY receive procedure of OFDM-based 802.11 protocols\cite{b12}. In the receiver end, the signal is first magnified by a LNA (Low Noise Amplifier) and filtered by a mixer. Then, the receiver uses STF (Short Training Field) and LTF (Long Training Field) to finish AGC (Automatic Gain Control), frequency correction and symbol timing adjust. Meanwhile, in the SIG (Signal Field), the MCS (Modulation and Coding Scheme) and length of the packet is also provided. \par
After that, the received sequence is transformed from serial to parallel, which is a procedure of accumulating sample point. Then follows the removement of GI (Guard Interval) of every OFDM Symbol, and FFT (Fast Fourier Transform) to frequency domain subcarrier for further processing.\par
After FFT, more detailed processing using LTF and corresponding pilot information to the OFDM Symbol is needed to remove FFO (Fractional Carrier Frequency Offset). This step includes channel equalizer, pilot tracking, phase correction and form some reliability information for decoder. The reason why pilot tracking is used is that it can correct the imperfections in the equalizer response, since that the equalizer response is not perfect.
Since we have changed the phase of the original signal, applying pilot tracking and phase correction will cause the loss of the MOXcatter tag information modulated on the signal. In order to retain the tag information, the pilot phase tracking must be switched off. Undoubtedly, this will cause the rising of the PER (Packet Error Rate) of the tagged signal, but still acceptable thanks to other frequency correction step in PHY receive procedure.\par
However, because of the absence of the pilot phase tracking, the FFO can't be remove properly, and the residual frequency offset causes phase distortion. As the phase distortion accumulates, the correctness of the decoded MOXcatter tag information is largely affected by the phase ambiguity. \par
To solve this problem, we change the modulation method from BPSK to DBPSK, and the comparison of these two methods is shown in Figure 4. It shows how the BER (Bit Error Rate) changes when SNR (Signal-Noise Ratio) changes.
\begin{figure}[htbp]
    \centerline{\includegraphics[width=0.9\linewidth]{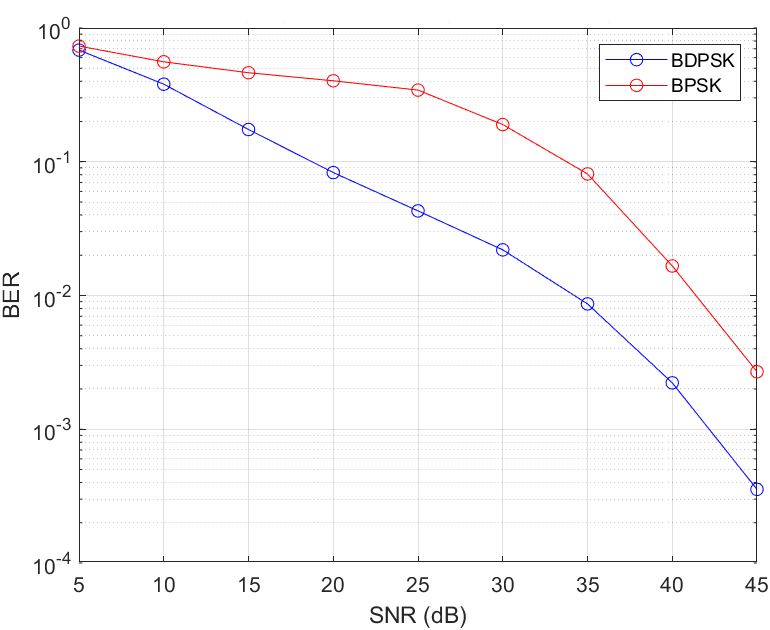}}
    \caption{Tag error bit rate changed with signal-noise ratio in different modulation methods}
    \label{fig4}
\end{figure}
The distance between the transmitter and receiver is 10m, and the index of MCS (Modulation and Coding Scheme) is 0, indicates that the specification for the WiFi excitation signal is 802.11n MIMO-OFDM Single-Stream (SS) BPSK-subcarrier 13Mbps. The channel bandwidth is 20MHz, and the length of the Physical layer convergence procedure (PLCP) service data unit (PSDU) length is 1000 bytes. The channel conditions are set as Model-B NLOS, representing the residential environment such as intra-room or room-to-room communication, and there is no large scale fading effect.

\section{Evaluation}
In this section, we evaluate the DBPSK modulated MOXcatter performance under different channel conditions, including modulation rate, delay profile model, channel bandwidth, PSDU length, SNR and communication distance. The evaluation is conducted in PHY, and the performance is examined through the BER.

\subsection{Experiment Setup}\label{AA}
We conduct our experiments in Matlab simulation. In a standard 802.11n packet error rate simulation for 1x1 TGn channel, we modulate the MOXcatter tag information on the TX signal and then compare the decoded tag sequence to the reference sequence. When a packet is not detected or the packet delay is out of the range of expected delays, any tag bit modulated on the packet will be labeled as an error, and that's why the BER sometimes is higher than 0.5. In this experiment, all the SNR is measured in RX.

\subsection{Modulation Rate}
Modulation rate indicates how many OFDM symbols are used to modulate one tag bit. In the classic MOXcatter system, this rate is 2 OFDM symbol/tag bit. We set the specification for the WiFi excitation signal as 802.11n MIMO-OFDM Single-Stream (SS) BPSK-subcarrier 13Mbps, and the distance between the transmitter and receiver is 10m. The channel bandwidth is 20MHz, and the PSDU length is 1000 bytes. Still, the channel conditions are set as Model-B NLOS. 
\begin{table*}[htbp]
    \caption{Delay Profile Model}
    \begin{center}
    \begin{tabular}{|c|c|c|c|}
    \cline{1-4} 
    \textbf{Model} & \textbf{RMS delay spread (ns)}& \textbf{Environment}& \textbf{Example} \\
    \hline
    A & 0& N/A& N/A \\
    \hline
    B & 15& Residential& Intra-room, room-to-room \\
    \hline
    C & 30& Residential/small office& Conference room,classroom \\
    \hline
    D & 50& Typical office& Sea of cubes, large conference room \\
    \hline
    E & 100& Large office& Multi-story office, campus small hotspot \\
    \hline
    F & 150& Large space (indoors/outdoors)& Large hotspot, industrial, city square \\
    \hline
    \end{tabular}
    \label{tab2}
    \end{center}
    \end{table*}
    \begin{figure}[htbp]
        \centerline{\includegraphics[width=0.9\linewidth]{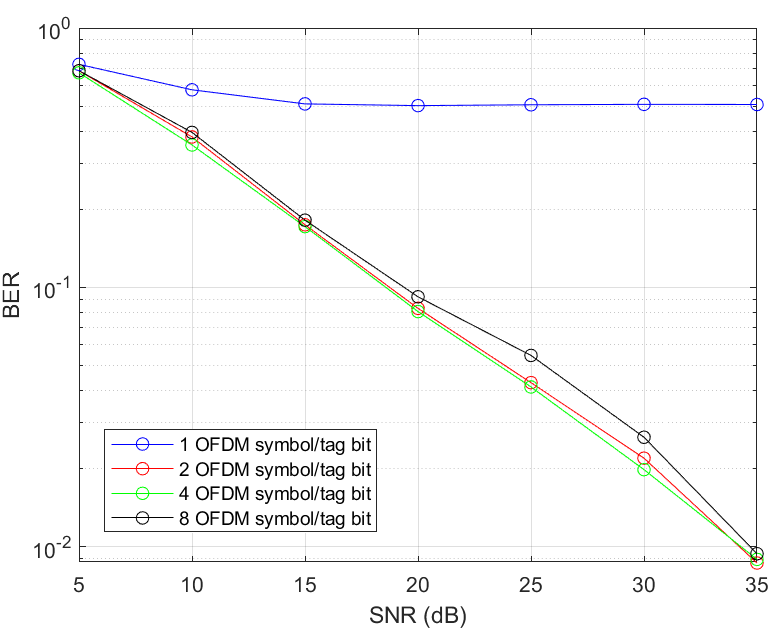}}
        \caption{Tag error bit rate changed with signal-noise ratio in different modulation rate experiments}
        \label{fig5}
    \end{figure}
We change the number of OFDM symbols used to modulate one tag bit from 1 to 8, as is shown in Figure 5.\par

When using only one OFDM symbol to modulate one tag bit, the tag information can't be recovered properly, which is also mentioned in []. This is because the bursty bit errors has more chance to cause tag bit errors. When we change the number to 2, the BER decrease sharply. 4 OFDM symbol/tag has almost the same performance as 2 OFDM symbol/tag bit, though the throughput is halved. However, keep increasing the number may leads to higher BER. As discussed before, fewer tag bit per packet will not always reduce the phase ambiguity. Though phase ambiguity has fewer chance to cause a tag bit error, it has a greater influence in BER since that the amount of phase ambiguity per packet stay still but the tag bit per packet is less. Obviously, using two OFDM symbols to modulate one tag bit is the rate that has the highest throughput.In the following experiments, we will keep using this modulation rate and other experiment setups if not mentioned.

    \begin{figure}[htbp]
        \centerline{\includegraphics[width=0.9\linewidth]{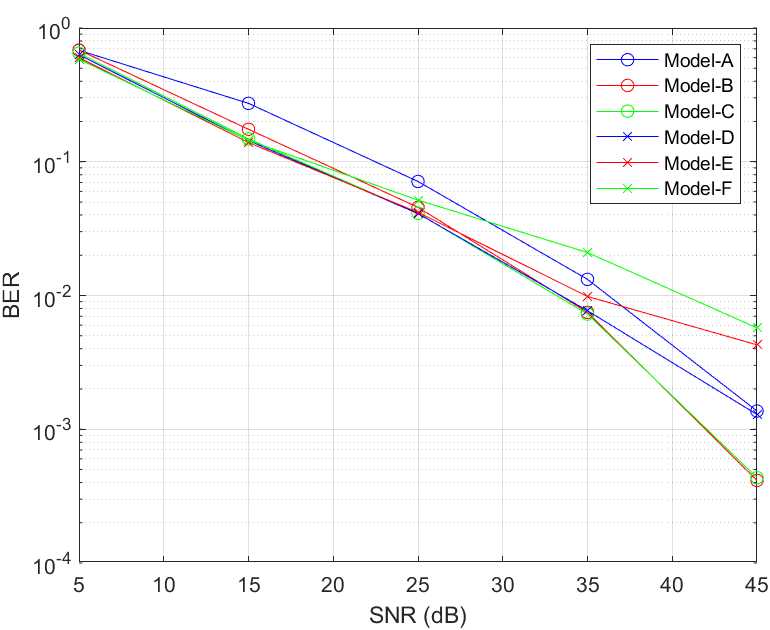}}
        \caption{Tag error bit rate changed with signal-noise ratio in different delay profile model experiments}
        \label{fig6}
    \end{figure}

\subsection{Delay Profile Model}
We then measure the BER in different delay profile models. In 802.11n protocol, there are 6 reference models, each representing a intra-room environment, as is shown in Table 2. In all 6 Models, there are no large scale fading effect since we've fixed the distance. Figure 6 shows the function of decoded tag data BER and signal-noise ratio in different delay profile models. Model-B and Model-C have almost the same performance and the lowest BER, and BER increases as the environment degrades in the order of Model-D, Model-E and Model-F. However, Model-A does not have the best performance though having the best condition, probably because the receiver is designed for multipath fading channel.

\subsection{Channel Bandwidth}
Channel bandwidth is also an important factor in MOXcatter system. In a Model-B NLOS channel, we measure the BER in 20MHz and 40MHz bandwidth and the result is shown in Figure 7. Unsurprisingly, higher bandwidth has lower BER. Compare 40MHz-2 OFDM symbol/tag bit to 20MHz-4 OFDM symbol/tag bit, these two modulation methods have the same sample point/tag bit and almost the same PSDU bit/tag bit (104 PSDU bit/tag bit in 20MHz-4 OFDM symbol/tag bit, 108 in 40MHz-2 OFDM symbol/tag bit). However, thanks to the lower FFO, the first one is obviously better than the latter.

\begin{figure}[htbp]
    \centerline{\includegraphics[width=0.9\linewidth]{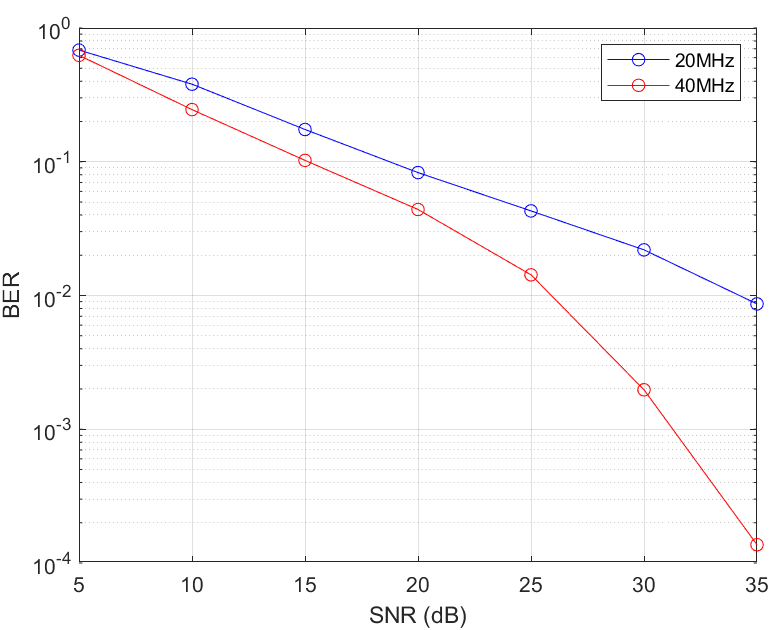}}
    \caption{Tag error bit rate changed with signal-noise ratio in different channel bandwidth}
    \label{fig7}
\end{figure}

\subsection{PSDU Length}
Another vital factor is PSDU length, indicating how many bytes are there in a PSDU packet. We change this length from 480 to 2040, and according to Figure 8, longer PSDU packet is related to higher BER. This is because the frequency correction is conducted every packet, and shorter PSDU packet is less possible to be reflected by FFO. 

\begin{figure}[htbp]
    \centerline{\includegraphics[width=0.9\linewidth]{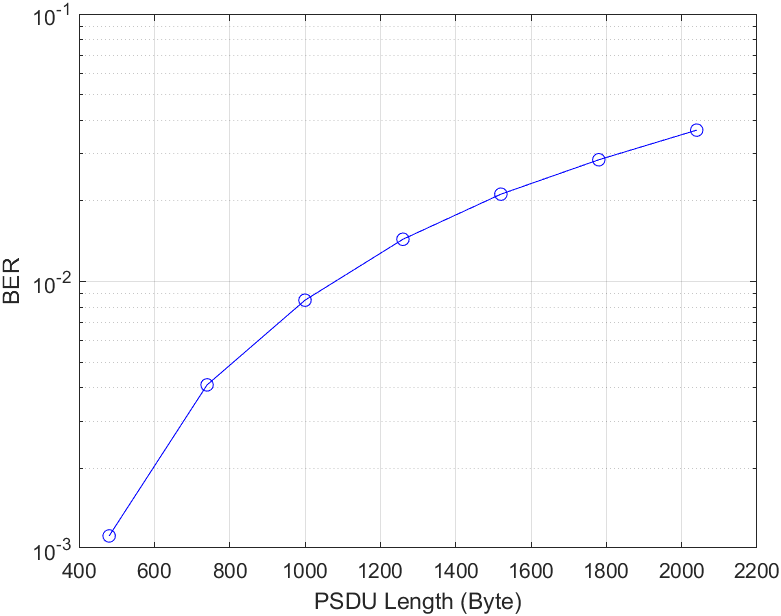}}
    \caption{Tag error bit rate changed with PSDU length}
    \label{fig8}
\end{figure}
However, shorter packet may leads to lower throughput of the original data communication since preamble sequence will not be shorten.
\subsection{Signal-Noise Ratio}
Although the experiments before are mainly based on SNR, there are still something not been discussed yet. We can see from Figure 9 that, when SNR is low, the BER is mainly decided by PER, because every packet error will lead to big increase in BER; and when SNR is high, PER is close to zero and BER fell sharply in the meanwhile. In other words, the backscattered signal's energy must be guaranteed so as to achieve better performance.

\subsection{Communication Distance}
Communication distance is also an important factor for many IoT devices, since its direct correlation with SNR. Assume that $\rm{SNR_A} (dB)$ is the SNR when the distance between transceiver and receiver is $\rm{A} (m)$ , the SNR in $\rm{B} (m)$ can be 

\begin{figure}[htbp]
    \centerline{\includegraphics[width=0.9\linewidth]{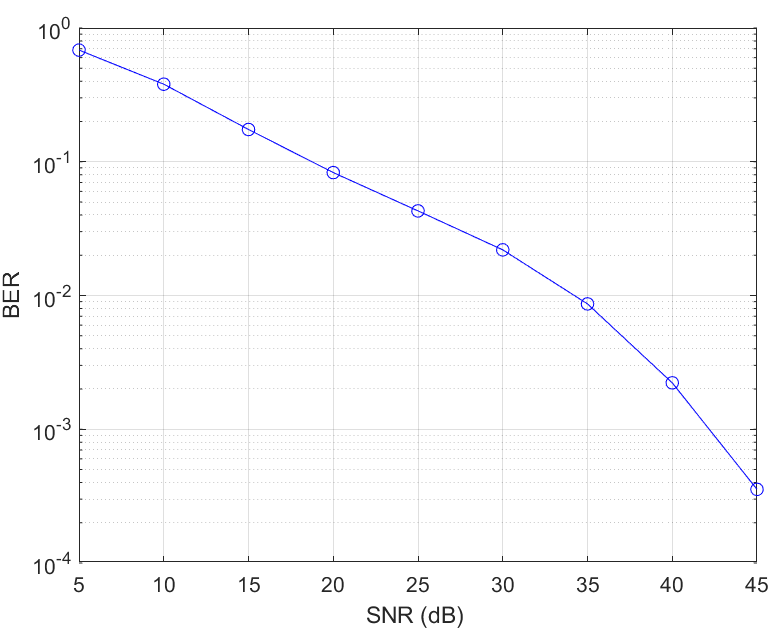}}
    \caption{Tag error bit rate changed with signal-noise ratio}
    \label{fig9}
\end{figure}

\begin{figure}[htbp]
    \centerline{\includegraphics[width=0.9\linewidth]{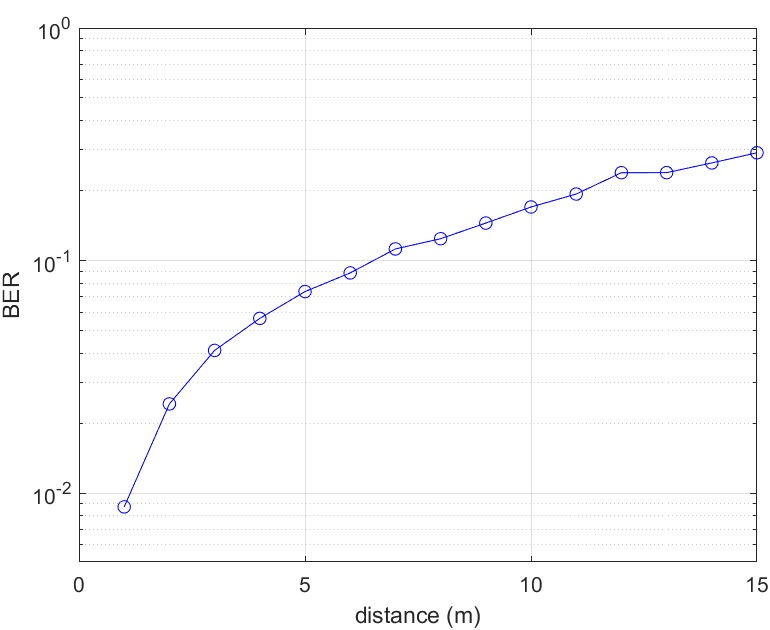}}
    \caption{Tag error bit rate changed with communication rate when excitation signal strength is fixed}
    \label{fig10}
\end{figure}
\noindent formulated as $\rm{SNR_A-20lg\frac{B}{A} (dB)}$ if the excitation signal strength is fixed. \par
We first fixed the excitation signal strength, while SNR in 1m is 35dB and 15dB in 10m. As is shwon in Figure 10, as the distance increases, the BER rises rapidly. Although the rises has slowed down in longer distance, its performance is not acceptable due to the poor SNR.\par
We then make the excitation signal strengthens while the distance between transceiver and receiver rises, keeping the SNR in the receiver 35dB in the whole process. Figure 11 shows that the BER almost doesn't change at all. Obviously, in our simulation environment, the most important factor is SNR. Multipath fading effect's influence to the system is stable in different distances.

\begin{figure}[htbp]
    \centerline{\includegraphics[width=0.9\linewidth]{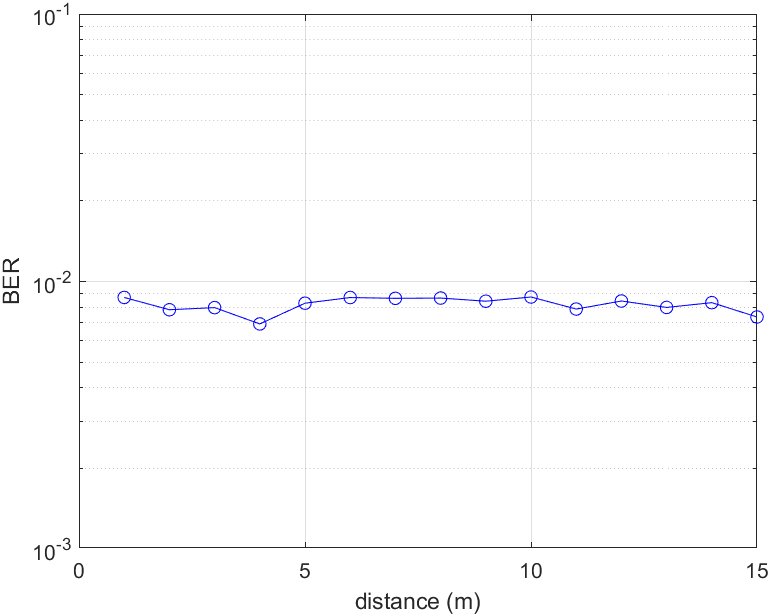}}
    \caption{Tag error bit rate changed with communication rate when signal-noise ratio is fixed}
    \label{fig11}
\end{figure}

\section{Application Case}
Obviously, the most common application of MOXcatter system is smart-home devices. Given that it's completely battery-free, we can assume that the distance between MOXcatter tag and the provider of the excitation signal should be limited to a small range to ensure the energy supply is sufficient, which means it won't be moved most of the time. Based on this assumption, there are many applications that MOXcatter can be the prominent choice.
\begin{itemize}
\item Thermometer and hygrometer. For such devices, whether having a battery that need to be charged occasionally or need to be plugged in doesn't seem to be a good solution, especially when we expect it to update frequently and control the temperature and humidity reversely through smart-home air-conditioner and humidifier. MOXcatter is undoubtedly one of the best choices in such situation.
\item Security service. Also based on the battery-free characteristic, we can make an alarm that rings when it can't receive MOXcatter's backscattered signal. It's an outstanding choice, because the alarm can't be turned off by shutting down the message sending device. And thanks to MOXcatter, the alarm doesn't need to turn off for battery changing.
\item Device trigger. To reduce energy consumption of some high energy-consumption facilities, MOXcatter can act the role of a trigger using sensors. For example, turn on the CCTV (Closed Circuit Television) when moving object detected by infrared detectors.
\end{itemize}
 
\section{Further Discussion}
\subsection{DPSK in Higher Order Modulation}
The problem of phase ambiguity due to FFO still exists in higher order modulation such as QPSK, 16-QAM and 64-QAM. Under these circumstances, we can change the tag modulation method from QPSK to DQPSK to achieve better performance. However, DPSK is not suitable in MIMO environment since there's only one tag bit in a packet.
\subsection{Partial Packet Modulation}
As is discussed in section 4.E, modulating tag bits on the whole packet suffer from phase ambiguity. Correcting the error bits needs more redundancy, which may lead to lower throughput. However, modulating tag bits only on the front part of the packet has higher BER performance. Here is a trade-off between more bits but with more redundancy and fewer bits but with fewer redundancy. For the highest throughput, there may be a best number of bits to modulate on a packet in different channel state. This solution also works in MIMO systems, since the accumulated FFO is not enough to cause phase change shortly after synchronization.

\section{Conclusion}
In this paper, we enhanced MOXcatter by applying DPSK instead of normal PSK, and evaluated its performance in different network environments and channel states. We illustrate MOXcatter's design and its feasibility of modulating signal on a OFDM symbol without affecting the data communication of the original purpose, explain why using DPSK and how can it enhance the performance of the system, simulate the process and give interpretation for some phenomenon, present some outlook for MOXcatter's application,  discuss some remain problems in MOXcatter and their possible solutions and enhancements. We will continue to improve the research based on MOXcatter, probably in higher order modulation, MIMO networks and other advanced WiFi standards.

\end{document}